# Cloud detection in Landsat-8 imagery in Google Earth Engine based on a deep convolutional neural network

Zhixiang Yin[a,b,c], Feng Ling[a]*, Giles M. Foody[d], Xinyan Li[a], and Yun Du[a]

[a]*Key Laboratory for Environment and Disaster Monitoring and Evaluation, Hubei, Innovation Academy for Precision Measurement Science and Technology, Chinese Academy of Sciences, Wuhan 430077, China;*

[b]*Anhui Province Key Laboratory of Wetland Ecosystem Protection and Restoration, Anhui University, Hefei, Anhui 230601, China;*

[c]*University of Chinese Academy of Sciences, Beijing 100049, China;*

[d]*School of Geography, University of Nottingham, Nottingham NG7 2RD, U.K.*

*corresponding author – lingf@whigg.ac.cn

# Cloud detection in Landsat-8 imagery in Google Earth Engine based on a deep convolutional neural network

Abstract: Google Earth Engine (GEE) provides a convenient platform for applications based on optical satellite imagery of large areas. With such data sets, the detection of cloud is often a necessary prerequisite step. Recently, deep learning-based cloud detection methods have shown their potential for cloud detection but they can only be applied locally, leading to inefficient data downloading time and storage problems. This letter proposes a method to directly perform cloud detection in Landsat-8 imagery in GEE based on deep learning (DeepGEE-CD). A deep convolutional neural network (DCNN) was first trained locally, and then the trained DCNN was deployed in the JavaScript client of GEE. An experiment was undertaken to validate the proposed method with a set of Landsat-8 images and the results show that DeepGEE-CD outperformed the widely used function of mask (Fmask) algorithm. The proposed DeepGEE-CD approach can accurately detect cloud in Landsat-8 imagery without downloading it, making it a promising method for routine cloud detection of Landsat-8 imagery in GEE.

## 1. Introduction

Remotely sensed images acquired by Landsat sensors are of considerable importance to a variety of applications including land cover mapping, environmental monitoring, and the estimation of land surface variables (Wulder et al. 2019). Recently, various applications based on Landsat imagery have been increasingly performed with Google Earth Engine (GEE) (Schwatke, Scherer, and Dettmering 2019; Long et al. 2019; Liu et al. 2020), which provides a free platform to acquire and analyze a potentially large mass of remotely sensed data conveniently (Shelestov et al. 2017; Gorelick et al. 2017). In many applications, cloud-free imagery is required and hence the detection of cloud in Landsat images is often a prerequisite (Wu et al. 2016; Baraldi and Tiede 2018). Cloud-contaminated imagery may be ignored as their inclusion could have negative impacts on the application (Zhu and Woodcock 2014).

A variety of methods could be used to detect cloud in Landsat images in GEE, such as the function of mask (Fmask) (Zhu and Woodcock 2012; Zhu, Wang, and Woodcock 2015), the multitemporal cloud detection method (Mateo-García et al. 2018). Fmask extracts clouds using rules which are determined by their distinct physical characteristics (Qiu, Zhu, and He 2019). Although Fmask is well known and has been widely used, it still has several limitations. First, it is difficult to design a physical rule which is appropriate for all conditions. For example, Fmask may perform poorly in terms of cloud detection in mountainous regions (Qiu, Zhu, and He 2019). Additionally, Fmask mainly focusses on low-level spectral features and neglects the spatial pattern of cloud which can lead to misidentification (Jeppesen et al. 2019). For the multitemporal cloud detection method, it might not perform well in situations with sudden changes in the underlying surface (Mateo-García et al. 2018).

Recently, deep convolutional neural network (DCNN) based methods have become popular for cloud detection (Xie et al. 2017; Drönner et al. 2018; Shao et al. 2018; Chai et al. 2019; Li et al. 2019; Wieland, Li and Martinis 2019; Mateo-García et al. 2020). Different to Fmask that designs physical rules manually, the DCNN-based methods directly learn high-level features from training data. As both spectral and spatial information of cloud are used in the DCNN-based methods, high cloud detection accuracies could be acquired. In practice, however, the implementation of these DCNN-based methods is inconvenient, as they typically run locally, which means that Landsat images must be downloaded before cloud detection, potentially resulting in substantial time wasting and large storage requirements for the data.

In GEE, many powerful application programming interfaces (APIs) are offered to produce flexible applications, which makes it possible to run a DCNN model in GEE

(Wang et al. 2020). The ability to run DCNN-based cloud detection methods directly in GEE, would make it possible to achieve accurate cloud detection without retrieving and downloading huge amount of data (Adepoju and Adelabu 2020). In this letter, an approach to integrate DCNN in GEE for cloud detection in Landsat-8 imagery, termed as DeepGEE-CD, is proposed. A DCNN cloud detection model is first trained locally, and then the trained model is implemented in GEE with the help of the provided APIs, making it is possible to directly detect cloud for Landsat-8 imagery in GEE.

## 2. Methods

### 2.1. The DeepGEE-CD framework

A two-step cloud detection framework was adopted (Figure 1). In the first step, a lightweight multi-level feature connected DCNN is trained with Landsat-8 images and corresponding cloud masks. In the second step, the trained network is deployed in the cloud infrastructure provided by GEE.

#### 2.1.1. DCNN model training

In the proposed DeepGEE-CD framework, an appropriate DCNN for cloud detection should be first trained. Generally, the more layer types provide the more opportunity for hierarchical re-composition of extracted features and thus better learn the spatial pattern of cloud. However, due to the limitation of computing resource of GEE, the provided APIs do not support user-defined network with arbitrary layers. Therefore, the structure of DCNN was designed according to the APIs provided by GEE.

Here, a lightweight multi-level feature connected DCNN was used (Figure 2). The DCNN contains two symmetrical parts, one for down-sampling and the other for up-sampling. The basic component of the two symmetrical parts is a stacked module containing <u>T</u>wo parts, each of which consists of a <u>C</u>onvolution layer comprising 64 filters, a <u>B</u>atch normalization layer and a <u>P</u>arametric rectified linear unit (PReLU) (Nair and Hinton 2010), termed here as TCBP. The filter sizes of the first convolution layer in the down-sampling part and of the first convolution layer in each TCBP of the up-sampling part is 3×3×10 and 3×3×128, respectively, while all other filters have size of 3×3×64. A max-pooling layer which contains 64 filters of size 2×2 is added behind each TCBP in the down-sampling procedure to amplify the receptive filed. Similarly, an up-sampling modules with a scaling factor of 2 is used before each TCBP in the up-sampling part to gradually recover the size of abstract feature maps. Additionally, concatenations are applied to connect multi-level feature maps. The last layer after the up-sampling part comprises a single filter which has size 3×3×64 and a softmax function as the activation function to estimate the class label (cloud or not cloud) of every pixel.

The DCNN model was trained locally using the Pytorch framework. The Adam algorithm was used as the gradient descent optimization method in the back-propagation to train the DCNN, and the learning rate of Adam was empirically set to 0.0001.

*2.1.2. Cloud detection in GEE*

Once the multi-level feature connected DCNN was trained, it was then deployed in GEE. All layers (e.g. convolution and max-pooling) of the DCNN model were implemented with the APIs provided by GEE, and the trained parameters were imported into the DCNN model in GEE for cloud detection.

In total, the proposed DCNN model includes seven types of layers: convolution, max-pooling, up-sampling, concatenation, batch normalization, activation function unit, and softmax. Their roles are:

(1) Convolution applies several filters sliding through the input feature maps to extract and integrate high-level features and is formed using convolutions operation API which support for kernels with user-defined parameters and sizes.

(2) Max-pooling reduces the size of input, and reprojection and reducing resolution APIs are combined to implement it.

(3) Up-sampling enlarges the size of input and is realized through reprojection and resampling APIs.

(4) Image cat API is used to concatenate feature maps with the same size in the down-sampling and up-sampling.

(5) Batch normalization actively centers and rescales each input back to a given mean and standard deviation for avoiding divergence problem of DCNN, PReLU aims to form the nonlinear relationship between input and output, and softmax converts the input to the probability of cloud. They only perform arithmetic operations on the input and are implemented by combining some specific arithmetic APIs, such as add and square root.

Additionally, the convolution layer, batch normalization layer and activation unit (PReLU) were integrated as a single function, which supports for user-defined parameters (e.g., number of input channels of a convolution), to avoid a large number of duplicated codes. Note that, unlike convolution provided by normal deep learning frameworks, which can be used for four-dimension tensors, convolutions in GEE are only support for two-dimension images. In this study, we combined arithmetic and cyclical mapping APIs to achieve four-dimension convolution. Each of max-pooling and up-sampling was also packaged as a function for convenient use.

Importing trained parameters into the DCNN is another necessary step for cloud detection in GEE. Originally, the trained parameters were saved in a binary model file.

They were converted to text data and stored in tables and up-loaded to GEE assets module in this study. Parameters were then extracted from the tables and assigned to corresponding variables. For example, weight parameters of each convolution layer were used to create kernels, the only argument to convolutions in GEE.

After deploying the DCNN and importing trained parameters in GEE, a cloud mask can be obtained online for a satellite image without downloading it.

*2.2. Comparator method and accuracy assessment*

The generated cloud maps from DeepGEE-CD were compared with those produced by Fmask and RS_Net (Jeppesen et al. 2019). In the function of Fmask, the output cloud pixels are dilated by a default value of 3 pixels (Qiu, Zhu, and He 2019), and this may lead to a high commission error. To compare these methods comprehensively, the produced cloud maps from Fmask without dilation were given as well. For accuracy assessment, the overall accuracy (OA), the commission error, the omission error, and the mean intersection over union (MIOU) were calculated using the reference cloud map (Tharwat 2018; Foody 2002). OA and MIOU were used to evaluate the overall performance, while the commission error and omission error were used in relation to the detection of cloud contaminated pixels. The most accurate result will have a high OA and MIOU as well as a low commission and omission error.

**3. Experiment**

*3.1. Data set and experimental setting*

The Landsat-8 cloud cover assessment validation data produced by the U.S. Geological Survey (USGS) Earth Resources Observation and Science (EROS) Center was used in this study (Foga et al. 2017, USGS. 2016). This data collection contains 96 Landsat-8

Operational Land Imager (OLI) and Thermal Infrared Sensor (TIRS) terrain-corrected (Level-1T) scenes with corresponding manually labeled cloud masks. Additionally, the data are evenly distributed over eight biomes (i.e., barren, forest, grass, shrubland, snow, urban, water, and wetland). In these scenes, each pixel is labeled as cloud, thin cloud, cloud shadow or clear. In this study, cloud and thin cloud were regarded as cloud, and cloud shadow and clear were regarded as non-cloud.

In the training process, 72 scenes were selected as the training data. All training data were cropped into image patches with size of 512 × 512 pixels without overlap, and a total of 9194 image patches, which were evenly distributed over the eight biomes, were generated. Ten bands of Landsat-8 imagery, all of which have pixel size of 30 m, were used as input. For spectral bands derived from OLI and TIRS, Top of Atmosphere (TOA) reflectance and Brightness Temperature (BT) were used, respectively. Batch size of training samples was set to 10.

In the cloud detection stage, DeepGEE-CD was run in the JavaScript client of GEE. The remainder 24 scenes of the Landsat-8 data set were used as the test data. Since the GEE provides a high-performance cloud computing platform and there are no strict requirements on input images' sizes for image operation in GEE, the input image of the deployed DCNN can be the entire Landsat-8 imaged scene and the DeepGEE-CD can output the corresponding cloud mask directly. Specifically, the test Landsat-8 imagery was imported from Landsat-8 collection 1 products which comprise TOA reflectance and BT, and it was then used as the input of DeepGEE-CD to produce corresponding cloud mask. Additionally, cloud mask produced by Fmask is provided as a layer in GEE and is used for comparative assessment.

### *3.2. Results and discussion*

To demonstrate the visual performance of the cloud detection methods, three examples are given in Figure 3. The results from Fmask with 3 pixels' dilation (Fmask_D3) overestimated many cloud pixels in all three cases since it dilated 3 pixels in eight directions. The results from Fmask without dilation (Fmask_D0) was better than those of Fmask_D3, but there still exist cloud overestimation in some areas (dashed red ellipse in Figure 3). In contrast, the deep learning-based RS_Net and DeepGEE-CD accurately identified most cloud, and results from it are much closer to the reference cloud mask.

Table 1 presents a summary of the quantitative evaluations of cloud detection according to different biomes as well as the overall performance for the used methods. Overall, the outputs from the RS_Net and proposed DeepGEE-CD were more accurate than those from Fmask_D3 and Fmask_D0. The overall OA, commission error, and MIOU for the proposed method are 0.96, 4.00%, and 0.90 respectively, which are similar to those from RS_Net, and are better than those associated with the use of Fmask_D3 and Fmask_D0. Additionally, the values of OA and MIOU obtained from the use of RS_Net and DeepGEE-CD show less inter-biome difference than Fmask_D3 and Fmask_D0. Note that, the result of DeepGEE-CD is comparable with RS_Net, which runs off-line and has more filters, indicating that the proposed lightweight DCNN is competent to form the nonlinear relationship between Landsat-8 imagery and corresponding cloud map.

The enhanced performance of DeepGEE-CD arises mainly from the ability of the DCNN to extract latent multi-level spatial/spectral features of the original Landsat-8 imagery and represent the spatial pattern of cloud from the training samples. In contrast, Fmask only uses low-level features according to designed rules, and some potential features may not be fully utilized. Note that, the DCNN-based model can also be implemented in GEE using TensorFlow framework based on Google AI platform,

Google cloud storage, and Google colaboratory. However, the AI platform is not free of charge, and the process of its computing environment configuration is complex. These factors may greatly limit its practicability in practical applications. In reality, GEE provides a high-performance cloud computing platform with a range of user-friendly APIs in JavaScript client library. This makes it possible to build the DCNN and import trained parameters in GEE. The ability of DeepGEE-CD to process a whole Landsat-8 scene at once makes DeepGEE-CD very convenient. Additionally, the uploaded tables in GEE assets module can be shared to other users, making DeepGEE-CD public available.

There are, however, two limitations for the proposed DeepGEE-CD. First, the activation unit does not support for customization in the packaged convolution function. As a result, the convolution function needs to be rewritten if a new activation unit is adopted. Second, limited by the computation resource of GEE, some specific convolution layers of DCNN cannot be implement in GEE. For example, dilated convolution layer could not be achieved due to the fact that dilation is not supported in the convolution API provided by GEE. Conversion other types of convolutions to the convolution used in this study may help to solve this problem and it needs further investigation.

## 4. Conclusion

This letter proposes the DeepGEE-CD that aims to employ deep convolutional neural network (DCNN) in GEE to achieve cloud detection for Landsat-8 imagery. Experimental results showed that the proposed method was effective, achieving highly accurate cloud detection in Landsat-8 imagery. DeepGEE-CD is an initial attempt for detecting cloud with deep learning-based model in GEE, there are some issues should be considered for further application. For example, cloud shadow was not considered in

DeepGEE-CD, and this can be solved by training a new DCNN model with sufficient cloud shadow samples including all different possibilities. It is also possible to extend the model to other satellite sensors and biome types by using a wide range of corresponding training data.

**Acknowledgments**

This work was supported in part by Hubei Provincial Natural Science Foundation for Innovation Groups (No. 2019CFA019), Strategic Priority Research Program of Chinese Academy of Sciences (No. XDA 2003030201), and the Natural Science Foundation of Anhui Province (No. 1908085QD161).

**Supplements**

The codes and uploaded data of DeepGEE-CD are available at: https://code.earthengine.google.com/94b0c6c75102a5952388762202f17388.

**References**

Adepoju, Kayode A., and Samuel A. Adelabu. 2020. "Improving accuracy of Landsat-8 OLI classification using image composite and multisource data with Google Earth Engine." *Remote Sensing Letters* 11 (2):107-16. doi: 10.1080/2150704X.2019.1690792.

Baraldi, Andrea, and Dirk Tiede. 2018. "AutoCloud+, a "Universal" Physical and Statistical Model-Based 2D Spatial Topology-Preserving Software for Cloud/Cloud–Shadow Detection in Multi-Sensor Single-Date Earth Observation Multi-Spectral Imagery—Part 1: Systematic ESA EO Level 2 Product Generation at the Ground Segment as Broad Context." *ISPRS International Journal of Geo-Information* 7 (12):457.

Chai, Dengfeng, Shawn Newsam, Hankui K. Zhang, Yifan Qiu, and Jingfeng Huang. 2019. "Cloud and cloud shadow detection in Landsat imagery based on deep convolutional neural networks." *Remote Sensing of Environment* 225:307-316. doi: 10.1016/j.rse.2019.03.007.

Drönner, Johannes, Nikolaus Korfhage, Sebastian Egli, Markus Mühling, Boris Thies, Jörg Bendix, Bernd Freisleben, and Bernhard Seeger. 2018. "Fast Cloud Segmentation Using Convolutional Neural Networks." *Remote Sensing* 10 (11):1782.doi: 10.3390/rs10111782.

Foga, Steve, Pat L. Scaramuzza, Song Guo, Zhe Zhu, Ronald D. Dilley, Tim Beckmann, Gail L. Schmidt, John L. Dwyer, M. Joseph Hughes, and Brady Laue. 2017. "Cloud detection algorithm comparison and validation for operational Landsat data products." *Remote Sensing of Environment* 194:379-90. doi: 10.1016/j.rse.2017.03.026.

Foody, Giles M. 2002. "Status of land cover classification accuracy assessment." *Remote Sensing of Environment* 80 (1):185-201. doi: 10.1016/S0034-4257(01)00295-4.

Gorelick, Noel, Matt Hancher, Mike Dixon, Simon Ilyushchenko, David Thau, and Rebecca Moore. 2017. "Google Earth Engine: Planetary-scale geospatial analysis for everyone." *Remote Sensing of Environment* 202:18-27. doi: 10.1016/j.rse.2017.06.031.

Jeppesen, Jacob Høxbroe, Rune Hylsberg Jacobsen, Fadil Inceoglu, and Thomas Skjødeberg Toftegaard. 2019. "A cloud detection algorithm for satellite imagery based on deep learning." *Remote Sensing of Environment* 229:247-259. doi: 10.1016/j.rse.2019.03.039.

Li, Zhiwei, Huanfeng Shen, Qing Cheng, Yuhao Liu, Shucheng You, and Zongyi He. 2019. "Deep learning based cloud detection for medium and high resolution remote sensing images of different sensors." *ISPRS Journal of Photogrammetry and Remote Sensing* 150:197-212. doi: 10.1016/j.isprsjprs.2019.02.017.

Liu, Luo, Xiangming Xiao, Yuanwei Qin, Jie Wang, Xinliang Xu, Yueming Hu, and Zhi Qiao. 2020. "Mapping cropping intensity in China using time series Landsat and Sentinel-2 images and Google Earth Engine." *Remote Sensing of Environment* 239:111624. doi: 10.1016/j.rse.2019.111624.

Long, Tengfei, Zhaoming Zhang, Guojin He, Weili Jiao, Chao Tang, Bingfang Wu, Xiaomei Zhang, Guizhou Wang, and Ranyu Yin. 2019. "30 m Resolution Global Annual Burned Area Mapping Based on Landsat Images and Google Earth Engine." *Remote Sensing* 11 (5):489. doi: 10.3390/rs11050489.

Mateo-García, Gonzalo, Luis Gómez-Chova, Julia Amorós-López, Jordi Muñoz-Marí, and Gustau Camps-Valls. 2018. "Multitemporal Cloud Masking in the Google

Earth Engine." *Remote Sensing* 10 (7):1079.doi: 10.3390/rs10071079.

Mateo-García, Gonzalo, Valero Laparra, Dan López-Puigdollers, and Luis Gómez-Chova. 2020. "Transferring deep learning models for cloud detection between Landsat-8 and Proba-V." *ISPRS Journal of Photogrammetry and Remote Sensing* 160:1-17. doi: /10.1016/j.isprsjprs.2019.11.024.

Nair, Vinod, and Geoffrey E. Hinton. 2010. "Rectified Linear Units Improve Restricted Boltzmann Machines". Paper presented at the international conference on machine learning (ICML), Haifa, June 21-24.

Qiu, Shi, Zhe Zhu, and Binbin He. 2019. "Fmask 4.0: Improved cloud and cloud shadow detection in Landsats 4–8 and Sentinel-2 imagery." *Remote Sensing of Environment* 231:111205. doi: 10.1016/j.rse.2019.05.024.

Schwatke, Christian, Daniel Scherer, and Denise Dettmering. 2019. "Automated Extraction of Consistent Time-Variable Water Surfaces of Lakes and Reservoirs Based on Landsat and Sentinel-2." *Remote Sensing* 11 (9):1010. doi: 10.3390/rs11091010.

Shao, Zhenfeng., Yin. Pan, Chunyuan. Diao, and Jiajun. Cai. 2019. "Cloud Detection in Remote Sensing Images Based on Multiscale Features-Convolutional Neural Network." *IEEE Transactions on Geoscience and Remote Sensing* 57 (6):4062-76. doi: 10.1109/TGRS.2018.2889677.

Shelestov, Andrii, Mykola Lavreniuk, Nataliia Kussul, Alexei Novikov, and Sergii Skakun. 2017. "Exploring Google Earth Engine Platform for Big Data Processing: Classification of Multi-Temporal Satellite Imagery for Crop Mapping." *Frontiers in Earth Science* 5 (17). doi: 10.3389/feart.2017.00017.

Tharwat, Alaa. 2018. "Classification assessment methods." *Applied Computing and Informatics*. doi: 10.1016/j.aci.2018.08.003.

U.S. Geological Survey. 2016. " Data from :L8 Biome Cloud Validation Masks" (dataset). U.S. Geological Survey. https://doi.org/10.5066/F7251GDH.

Wang, Y., Z. Li, C. Zeng, G. Xia, and H. Shen. 2020. "An Urban Water Extraction Method Combining Deep Learning and Google Earth Engine." *IEEE Journal of Selected Topics in Applied Earth Observations and Remote Sensing* 13:768-781. doi: 10.1109/JSTARS.2020.2971783.

Wieland, Marc, Yu Li, and Sandro Martinis. 2019. "Multi-sensor cloud and cloud shadow segmentation with a convolutional neural network." *Remote Sensing of Environment* 230:111203. doi: 10.1016/j.rse.2019.05.022.

Wu, Teng, Xiangyun Hu, Yong Zhang, Lulin Zhang, Pengjie Tao, and Luping Lu. 2016. "Automatic cloud detection for high resolution satellite stereo images and its application in terrain extraction." *ISPRS Journal of Photogrammetry and Remote Sensing* 121:143-156. doi: 10.1016/j.isprsjprs.2016.09.006.

Wulder, Michael A., Thomas R. Loveland, David P. Roy, Christopher J. Crawford, Jeffrey G. Masek, Curtis E. Woodcock, Richard G. Allen, et al. 2019. "Current status of Landsat program, science, and applications." *Remote Sensing of Environment* 225:127-147. doi: 10.1016/j.rse.2019.02.015.

Xie, F., M. Shi, Z. Shi, J. Yin, and D. Zhao. 2017. "Multilevel Cloud Detection in Remote Sensing Images Based on Deep Learning." *IEEE Journal of Selected Topics in Applied Earth Observations and Remote Sensing* 10 (8):3631-3640. doi: 10.1109/JSTARS.2017.2686488.

Zhu, Zhe, Shixiong Wang, and Curtis E. Woodcock. 2015. "Improvement and expansion of the Fmask algorithm: cloud, cloud shadow, and snow detection for Landsats 4–7, 8, and Sentinel 2 images." *Remote Sensing of Environment* 159:269-277. doi: 10.1016/j.rse.2014.12.014.

Zhu, Zhe, and Curtis E. Woodcock. 2012. "Object-based cloud and cloud shadow detection in Landsat imagery." *Remote Sensing of Environment* 118:83-94. doi: 10.1016/j.rse.2011.10.028.

Zhu, Zhe, and Curtis E. Woodcock. 2014. "Automated cloud, cloud shadow, and snow detection in multitemporal Landsat data: An algorithm designed specifically for monitoring land cover change." *Remote Sensing of Environment* 152:217-234. doi: 10.1016/j.rse.2014.06.012.

Table 1. Cloud detection accuracy for Fmask_D3/Fmask_D0/RS_Net/DeepGEE-CD.
(Most accurate result highlighted in bold).

| Land-cover class | OA | Commission error (%) | Omission error (%) | MIOU |
|---|---|---|---|---|
| Barren | 0.91/0.93/**0.96**/0.95 | 21.90/14.47/7.28/**5.38** | **2.28**/3.34/2.55/4.04 | 0.81/0.85/**0.91**/**0.91** |
| Forest | 0.86/0.94/**0.96**/0.95 | 36.44/13.08/**2.84**/3.10 | **1.27**/2.79/5.11/6.13 | 0.72/0.87/**0.91**/0.89 |
| Grass | 0.92/0.95/**0.99**/0.98 | 36.17/21.60/**2.50**/4.13 | **0.26**/0.83/1.02/1.65 | 0.77/0.85/**0.96**/0.94 |
| Shrubland | 0.93/0.93/**0.93**/**0.93** | 17.45/7.11/3.18/**2.64** | 3.34/7.14/**8.45**/9.24 | 0.83/**0.85**/**0.85**/0.84 |
| Snow | 0.84/0.83/0.88/**0.90** | 25.58/23.99/15.77/**7.12** | 12.20/14.56/**10.49**/10.79 | 0.68/0.67/0.75/**0.80** |
| Urban | 0.88/0.93/**0.97**/**0.97** | 39.82/20.36/6.47/**3.72** | **0.17**/0.81/1.33/2.85 | 0.73/0.85/**0.93**/**0.93** |
| Water | 0.90/0.93/**0.94**/**0.94** | 18.67/11.18/9.35/**6.44** | **1.67**/2.87/2.43/4.83 | 0.82/0.87/**0.89**/**0.89** |
| Wetland | 0.96/0.97/**0.99**/0.98 | 8.68/5.04/**0.65**/0.66 | **0.28**/0.51/2.00/2.48 | 0.92/0.95/**0.97**/**0.97** |
| Overall | 0.90/0.93/0.95/**0.96** | 23.7/13.48/5.89/**4.00** | **2.81**/4.28/4.29/5.34 | 0.79/0.85/**0.90**/**0.90** |

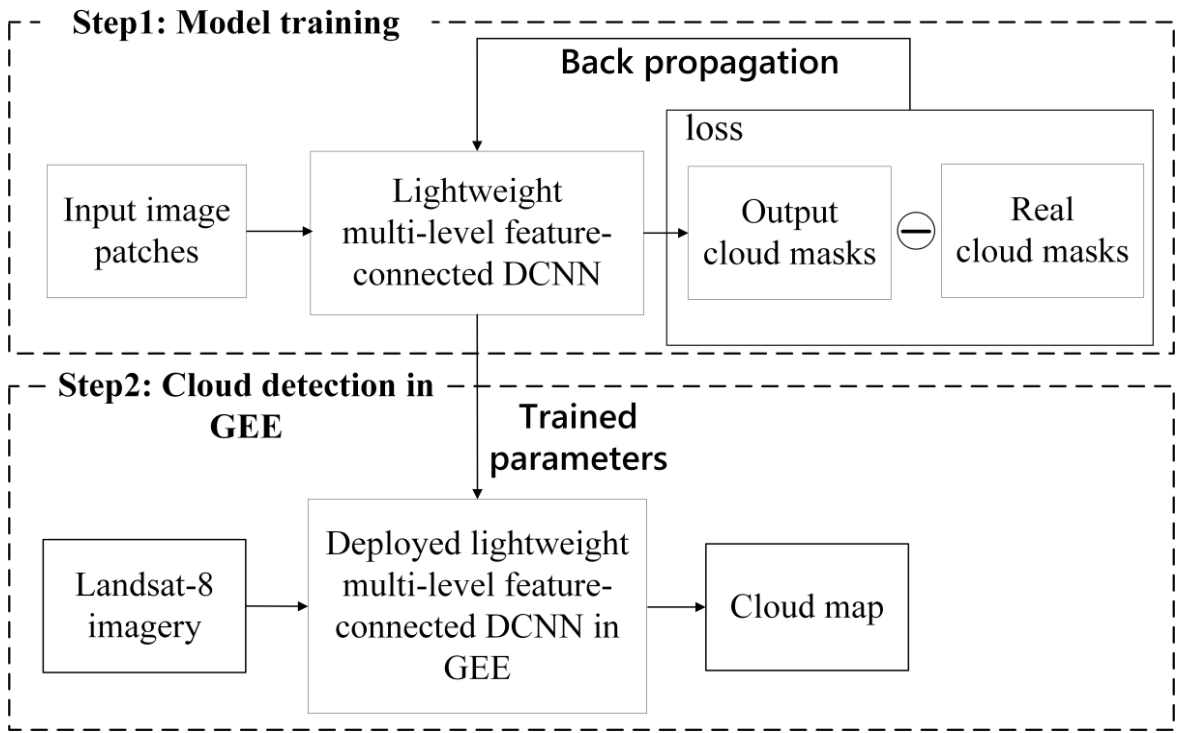

Figure 1. The flowchart of the proposed DeepGEE-CD framework.

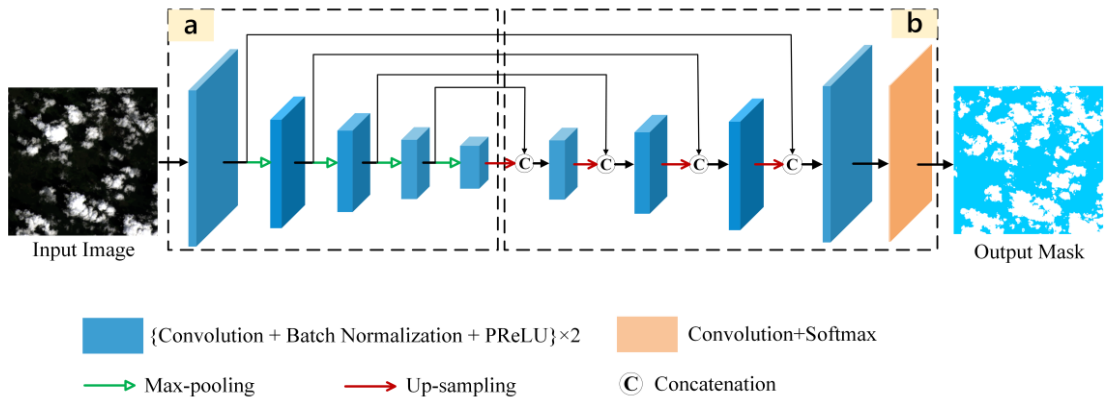

Figure 2. The architecture of the lightweight multi-level feature connected deep convolutional neural network (DCNN). (a) is the down-sampling part, and (b) is the up-sampling part.

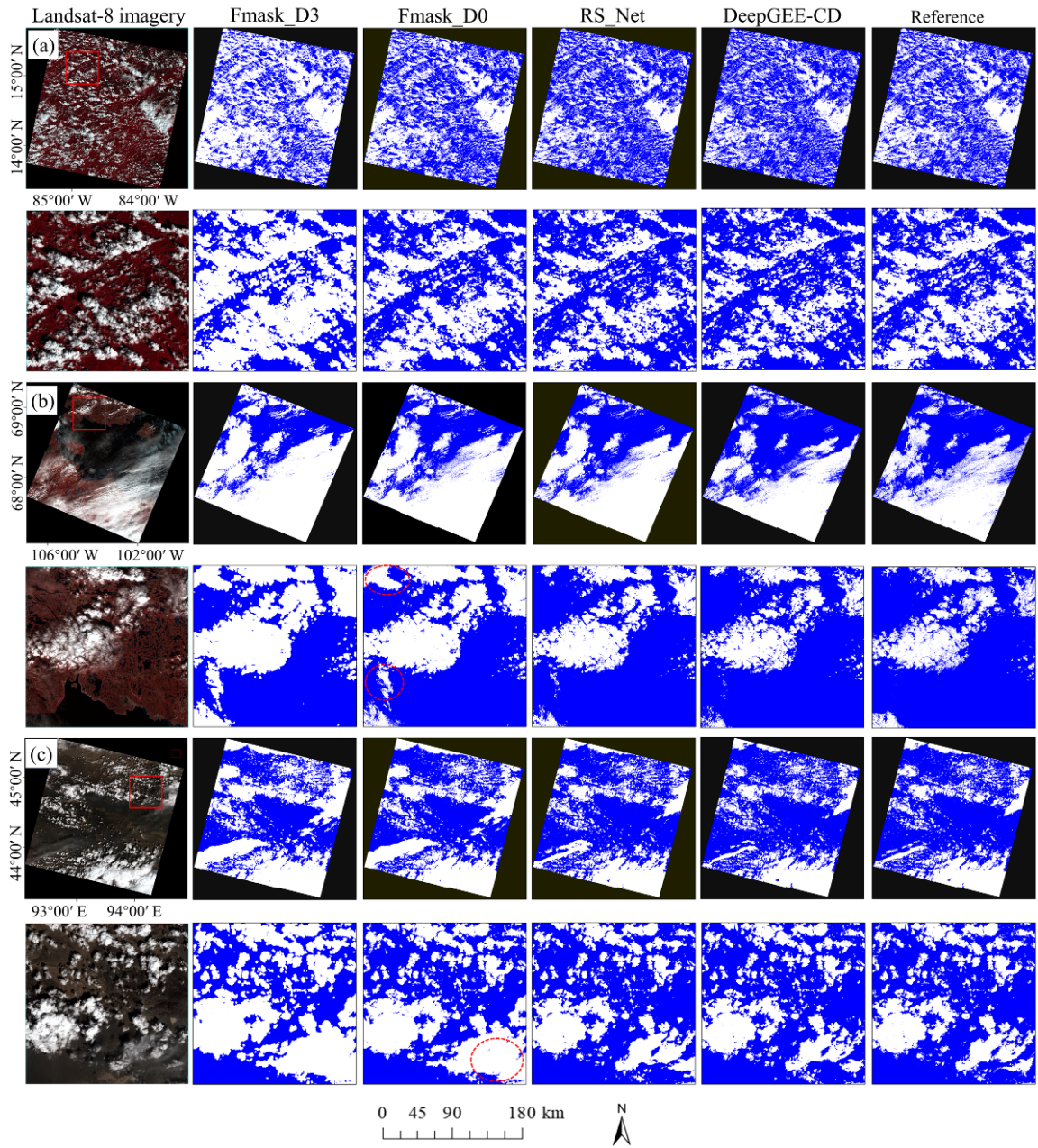

Figure 3. Examples of cloud detection results by Fmask_D3, Fmask_D0, RS_Net and DeepGEE-CD. Cloud and clear area are marked as white and blue, respectively. (a) was scene LC8_ p016r050_20140210, (b) was scene LC8_ p043r012_20140802, and (c) was scene LC8_ p139r029_20140515. The first row of each part shows the results of the whole scene, the second row depicts the zoomed-in figures of areas marked with red squares in the first row.